\documentclass[pdflatex,sn-mathphys-num,Numbered]{sn-jnl}

\usepackage{graphicx}%
\usepackage{multirow}%
\usepackage{amsmath,amssymb,amsfonts}%
\usepackage{amsthm}%
\usepackage{mathrsfs}%
\usepackage[title]{appendix}%
\usepackage{xcolor}%
\usepackage{textcomp}%
\usepackage{manyfoot}%
\usepackage{booktabs}%
\usepackage{array}%
\usepackage{algorithm}%
\usepackage{algorithmicx}%
\usepackage{algpseudocode}%
\usepackage{listings}%
\usepackage{pdfpages}%
\usepackage{subfig}%
\usepackage{setspace}%
\newcommand{\rev}[1]{{\color{black}#1}}

\begin{document}

\title{\rev{Korean Space Collision Environment Assessment Framework Based on 3D-Cell Model}}

\author[1]{\fnm{Jaewoo} \sur{Kim}}\email{jw.kim@kaist.ac.kr}
\author[1]{\fnm{Minchan}\sur{Song}}\email{smc0109@kaist.ac.kr}
\author[2]{\fnm{Jinsung}\sur{Lee}}\email{jinsung.lee@khu.ac.kr}
\author[3]{\fnm{Jiwoong}\sur{Yu}}\email{jiwoongyu@kasi.re.kr}
\author[3]{\fnm{Hosik}\sur{Kam}}\email{kam@kasi.re.kr}
\author[3]{\fnm{Jung Hyun}\sur{Jo}}\email{jhjo39@kasi.re.kr}
\author[3,4]{\fnm{Eun Jung}\sur{Choi}}\email{eunjung@kasi.re.kr}
\author[3,4]{\fnm{Jin}\sur{Choi}}\email{rutcome@kasi.re.kr}
\author*[1]{\fnm{Jaemyung} \sur{Ahn}}\email{jaemyung.ahn@kaist.ac.kr}

\affil[1]{\orgdiv{Department of Aerospace Engineering}, \orgname{Korea Advanced Institute of Science and Technology}, \orgaddress{\street{291 Daehak-ro}, \city{Daejeon}, \postcode{34141}, \state{Yuseong-gu}, \country{Korea}}}
\affil[2]{\orgdiv{Department of Astronomy and Space Science}, \orgname{Kyung Hee University}, \orgaddress{\street{1732 Deogyeong-daero}, \city{\rev{Yongin-si}}, \postcode{17104}, \state{Giheung-gu}, \country{Korea}}}
\affil[3]{\orgname{Korea Astronomy and Space Science Institute}, \orgaddress{\street{776 Daedeokdae-ro}, \city{Daejeon}, \postcode{34055}, \state{Yuseong-gu}, \country{Korea}}}
\affil[4]{\orgdiv{Astronomy and Space Science}, \orgname{University of Science and Technology}, \orgaddress{\street{217 Gajeong-ro}, \city{Daejeon}, \postcode{34113}, \state{Yuseong-gu}, \country{Korea}}}

\doublespacing
\abstract{\rev{Space situational awareness (SSA) requires purpose-matched models across spatial, temporal, and fidelity scales. Building on our previously reported three-dimensional (3D) cell formulation and implementation, this study establishes a reproducible, resolution-aware, catalog-conditioned framework for macroscopic assessment of the low Earth orbit (LEO) collision environment. The framework maps supplied catalog or scenario populations to time-averaged spatial density and target-specific impact metrics while retaining individual-object information. Using a 2025 Space-Track snapshot, we evaluate radial, declination, and right-ascension resolution sensitivity and computational performance for six targets, including two Korean space assets. Normalized expected impact counts range from 0.615 to 1.599 and vary nonmonotonically; for a synthetic 500-km circular target, the result at a 0.25-km radial width is 38.5\% below the 10-km reference. Runtime and memory show direction-dependent trade-offs. Ten annual snapshots show catalog growth from 15,723 objects in 2016 to 28,540 in 2025 and a 7.86-fold increase in the 500-km target metric, driven primarily by Starlink, other payloads, and unknown/TBA records. In a conditional stress test of the proposed 998,240-satellite SpaceX Orbital Data Center population, exact annual probabilities of at least one impact reach $3.75\times10^{-3}$ and $1.48\times10^{-3}$ for the 700-km and 1,000-km targets. The framework provides a reproducible, resolution-aware basis for catalog-conditioned environment monitoring, comparative scenario assessment, and prioritization of cases for higher-fidelity follow-up analysis.}}


\maketitle

\section{Introduction}\label{sec:introduction}

Recent technological advances---including the reduction in launch costs driven by increased capacity, reusable vehicles, and ride-sharing services, alongside lower manufacturing costs due to electronics miniaturization and COTS components---have made space more accessible and intensified various space activities. Consequently, the orbital environment around the Earth is becoming increasingly congested. Beyond the surge in satellite deployments, fragmentation events~\cite{esa_fragmentation_database}, including collisions~\cite{iridium_cosmos_collision}, explosions~\cite{intelsat33e_explosion}, and deliberate acts~\cite{chinese_asat, russian_asat}, are significant drivers of this increased object density. According to ESA's Annual Space Environment Report 2025~\cite{esa2025annual}, Earth orbit contains 54,000 objects greater than 10 cm (including approximately 9,300 active payloads), 1.2 million objects between 1 cm and 10 cm, and 130 million objects from 1 mm to 1 cm. This growing population \rev{increases collision risk to valuable space assets and, in the worst case, could trigger the catastrophic cascade known as the Kessler Syndrome~\cite{kessler1978collision, kessler2010kessler}.} Therefore, space situational awareness (SSA)---the ability to track, predict, and characterize the behavior of space objects---has become essential for \rev{stakeholders} to ensure the safety of their assets, secure the sustainability of space utilization, and contribute to the development of global regulatory measures.

\rev{SSA decisions range from operational responses to individual conjunctions to policy-level assessments of the evolving orbital environment. Supporting this range of decisions requires models tailored to their specific purposes, with appropriate spatial and temporal scales, population representations, and levels of modeling detail. Accordingly, organizations have developed tools ranging from fragmentation models to integrated frameworks for strategic planning and operations. Notable examples include efforts by NASA  and MIT in the United States~\cite{osma_od,matney2023overview,liou2004legend,dambrosio2023novel,jang2025new}; ESA in Europe~\cite{esa_sdup,wiedemann2020improvements,martin2004introducing,martin2004space,lewis2001damage}; CNSA in China~\cite{wang2019introduction}; CNES in France~\cite{dolado2013introducing}; the Institute of Space Systems at Technische Universit"{a}t Braunschweig in Germany~\cite{radtke2017luca2}; Kyushu University and JAXA in Japan~\cite{hanada2013orbital}; and ISTI-CNR in Italy~\cite{rossi2009new}.

Within this international context, the Korea Advanced Institute of Science and Technology (KAIST) and the Korea Astronomy and Space Science Institute (KASI) have been developing an integrated collision-risk analysis framework to support Korea's SSA capabilities~\cite{kim2024development,lee2024review}. Figure~\ref{fig:integrated_context} presents this framework, consisting of complementary microscopic and macroscopic analysis scales defined according to the analysis target, spatial granularity, and supported decision level. The microscopic, event-specific module uses the states and covariances of an asset and a risk object at the time of closest approach (TCA) to support decision-making for individual conjunction events. The macroscopic is for environment-level analysis, composed of two models with distinct purposes: the 3D-cell model and the 1D source-sink model. The 3D-cell model is for analysis of a snapshot of orbital status, including where traffic is concentrated or how much risk specific space assets are exposed. On the other hand, the 1D source-sink model is for the analysis of the evolution of the collision environment, with varying high-level drivers: for example, regulations on post-mission disposal. Among them, the current focus of this study, the 3D-cell model, is highlighted by the red box in Fig.~\ref{fig:integrated_context}.

\begin{figure}[hbt!]
    \centering
    \includegraphics[page=1,width=0.98\linewidth]{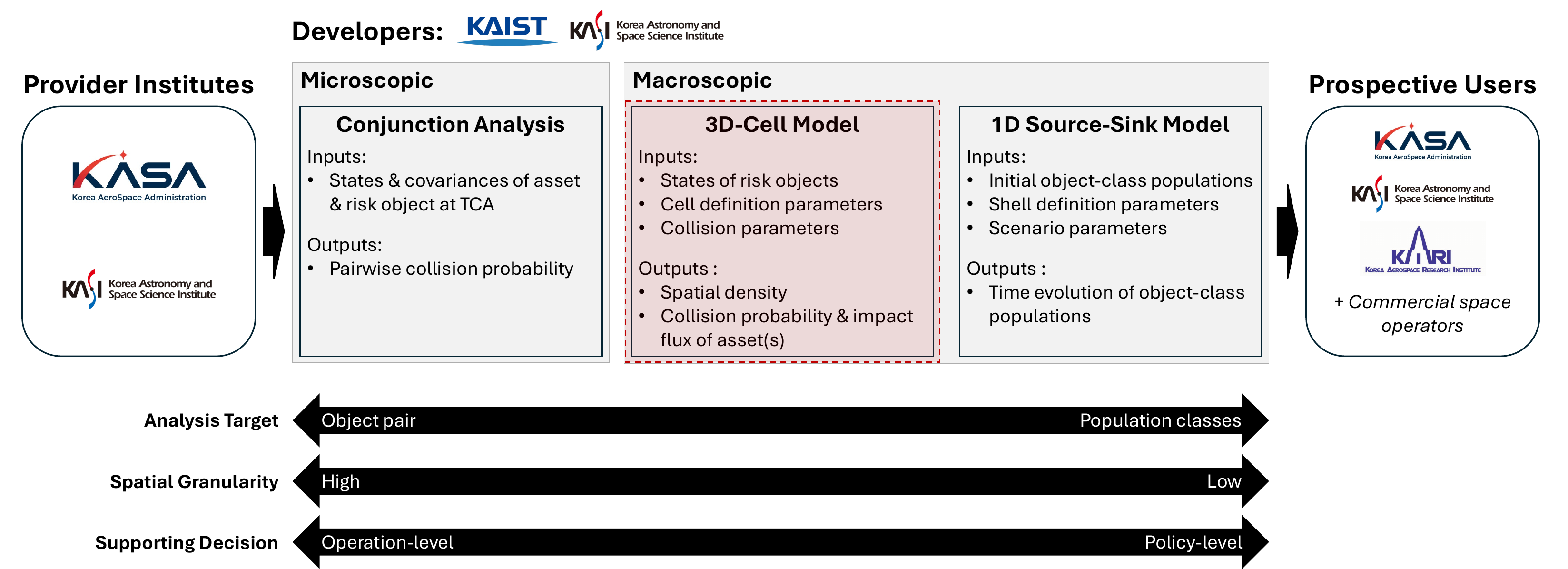}
    \caption{\rev{Korean integrated collision-risk analysis framework involving provider institutes, developers, and prospective users.}}
    \label{fig:integrated_context}
\end{figure}

The 3D-cell model discretizes the orbital environment in geocentric radius, declination, and right ascension. Catalogued-object states are mapped to these spatial analysis units, and collision flux is evaluated for specified target orbits based on the kinetic theory of gases~\cite{klinkrad2006space}, with the characteristics of individual catalogued objects used as inputs. Details of the formulation are provided later in this paper. This intermediate-detail representation is intended for rapid, reproducible macroscopic screening, with cases requiring greater fidelity directed to refined grids or microscopic analyses. The generic 3D-cell formulation, its initial implementation, and a comparison with MASTER-8.0.5 based on distinct population databases for three KITSAT-like Korean target orbits were presented in our previous study~\cite{kim2025development}.

Building on this foundation, the contribution of this paper is twofold. First, it establishes a reproducible, resolution-aware, catalog-conditioned 3D-cell workflow and quantifies the dependence of risk metrics and computational cost on spatial resolution. Second, it demonstrates the analytical use of this workflow through historical and future-scenario analyses of the orbital environment for representative target orbits, including orbits near those of Korean space assets.

The remainder of this paper is organized as follows. Section~\ref{sec:scope} positions the workflow relative to established model classes and the prior implementation and defines its intended use. Section~\ref{sec:method} introduces the mathematical foundations of the 3D-cell model. Section~\ref{sec:numerical_study} describes the experimental setups used throughout this study and presents sensitivity analyses of the cell-definition parameters across varying spatial resolutions. Section~\ref{sec:case_study} presents and discusses historical and future-scenario case studies using the same catalog and targets. Section~\ref{sec:conclusions} concludes the paper.

\section{Model Context and Application Scope}\label{sec:scope}

This section positions the current 3D-cell model relative to engineering-environment models, evolutionary population models, and component-level cell-based calculations. It then summarizes the previous comparison with MASTER and outlines the intended applications and users of the present workflow.

\subsection{Comparison to Existing Models}

Table \ref{tab:model_comparison} compares the key characteristics of representative engineering-environment models (ORDEM and MASTER), evolutionary population models (LEGEND and MOCAT), and the present 3D-cell workflow.

\begin{table}[hbt!]
\rev{
\caption{\rev{Key characteristics of representative approaches to collision-risk environment analysis.}}
\label{tab:model_comparison}
\centering
\small

{\setlength{\tabcolsep}{2.5pt}\renewcommand{\arraystretch}{1.15}%
\begin{tabular}{@{}>{\raggedright\arraybackslash}m{2.0cm}>{\centering\arraybackslash}m{2.65cm}>{\centering\arraybackslash}m{2.2cm}>{\centering\arraybackslash}m{2.25cm}>{\centering\arraybackslash}m{3.2cm}@{}}
\toprule
\textbf{Approach} & \textbf{Purpose} & \textbf{Population} & \textbf{Time basis} & \textbf{Output} \\
\midrule
ORDEM / MASTER & Environment characterization & Modeled population & Reference epochs & Flux; spatial density (MASTER) \\
LEGEND / MOCAT & Population evolution & Simulated population & Simulated evolution & Population projection \\
Present 3D-cell & Catalog-conditioned assessment & User-supplied population & Independent snapshots & Flux; spatial density \\
\botrule
\end{tabular}
}
}
\end{table}

ORDEM and MASTER characterize engineering environments using modeled populations that include sub-catalog particles. ORDEM provides target-specific impact fluxes, whereas MASTER provides both target-orbit flux and spatial-density spectra for specified environment epochs~\cite{nasa_ordem32,nasa_ordem32_userguide,esa_sdup,esa2020master_manual,braun2021master}. In contrast, LEGEND and the MOCAT family evolve scenario populations over time. MOCAT-SSEM aggregates species within altitude shells, whereas LEGEND and MOCAT-MC retain individual objects in higher-fidelity evolutionary simulations~\cite{liou2004legend,dambrosio2023novel,jang2025new}. These model classes address different temporal scales, levels of fidelity, and decision-making needs.

The present 3D-cell workflow serves a different role. It does not construct a complete engineering-environment population or propagate an orbital population through long-term evolution. Instead, it begins with an externally supplied catalog snapshot or a user-defined scenario population and evaluates the resulting environment at the analysis epoch. Catalog construction and epoch-state preparation are treated as upstream activities, allowing the analysis module to be applied to updated catalogs or alternative scenario populations without requiring a complete debris-source or population-evolution model.

Figure~\ref{fig:framework} summarizes this workflow. The supplied population and target definitions are processed through the 3D-cell analysis kernel to obtain time-averaged spatial density and target flux. Details of the individual modules are provided in Section~\ref{sec:method}. This separation between population preparation and environment analysis enables rapid, screening-level, catalog-conditioned comparative assessments, while the scope and completeness of the resulting products remain governed by the supplied population.

\begin{figure}[hbt!]
\centering
\includegraphics[page=2,width=0.98\linewidth]{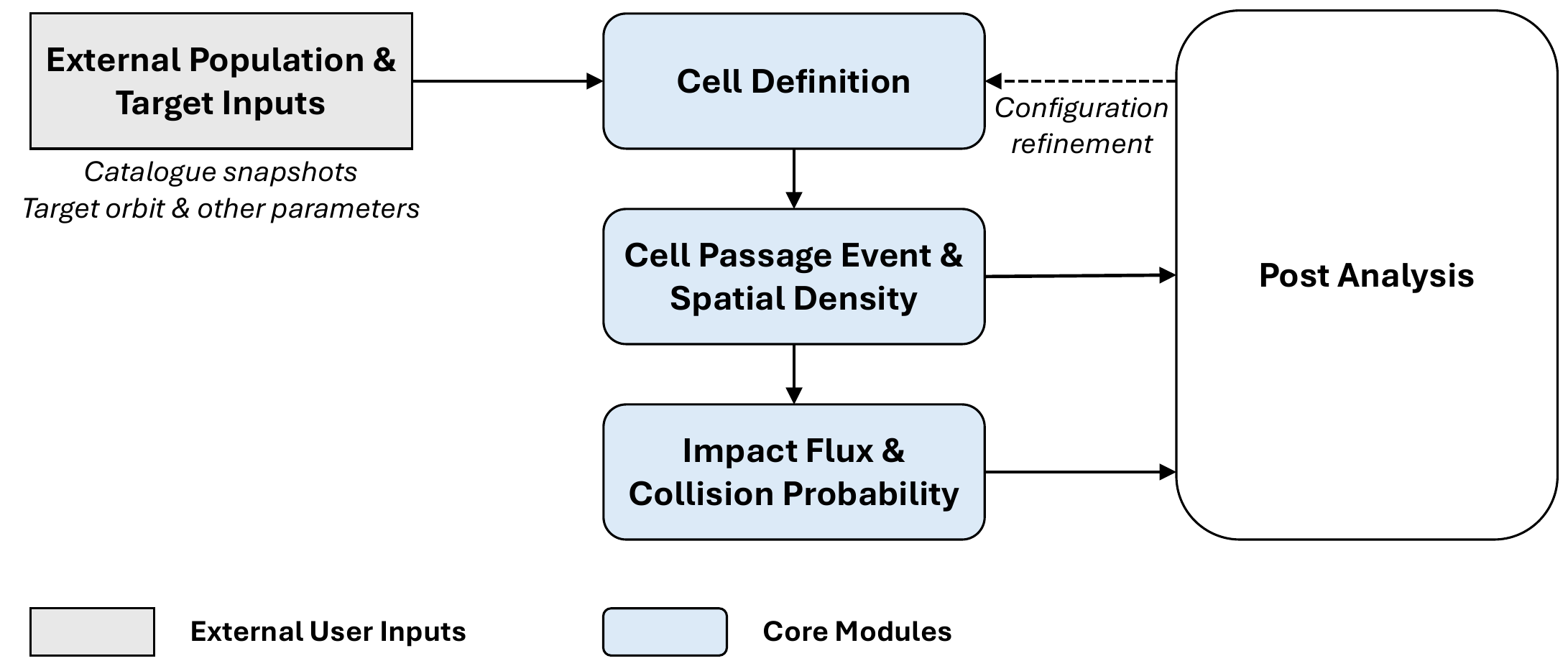}
\caption{\rev{Workflow of the 3D-cell analysis (gray blocks: external inputs; blue blocks: functional modules).}}
\label{fig:framework}
\end{figure}

The use of spatial cells for orbital-environment analysis also has precedents in established debris models, although the role of the cells and the associated calculations differ among approaches. MASTER transforms its modeled population into a 3D cell-passage-event (CPE) grid and derives target-orbit flux from target residence, particle density, and CPE-based target--impactor relative velocities~\cite{esa2020master_report,esa2020master_manual,klinkrad2006space}. Within LEGEND, Cube identifies potential collision pairs among objects occupying the same Cartesian cube at repeated simulation snapshots~\cite{liou2004cube}. These methods provide structural and algorithmic precedents for cell-based treatment of the orbital environment. The present 3D-cell workflow differs primarily in its intended role and implementation: it uses a supplied object population to construct a time-averaged spatial representation for macroscopic collision-flux assessment, rather than generating an engineering-environment population or identifying discrete collision events during long-term population evolution.

The generic 3D-cell formulation and its initial software implementation were reported in our previous study~\cite{kim2025development}. MASTER was selected as a particularly relevant cross-model reference because both approaches transform object populations into 3D cell-based representations and evaluate collision flux along specified target orbits~\cite{esa2020master_report,klinkrad2006space}. Using the August 2024 environment setting considered in the previous study, the 3D-cell model and MASTER-8.0.5 were evaluated for the same three KITSAT-like target orbits. The underlying populations were not identical: MASTER generated its environment from its internal population and source models, whereas the 3D-cell model used catalogued Space-Track states. However, the MASTER population was restricted to Launch- and Mission-Related Objects (LMRO), which served as a proxy for the catalogued-object population. The cell-definition parameters were also matched between the two approaches. The comparison was therefore intended as a cross-model consistency assessment rather than a point-by-point validation.

Despite these differences, the two models exhibited consistent altitude-dependent trends in spatial density, and the resulting target-orbit metrics were of the same order of magnitude, with 3D-cell-to-MASTER ratios ranging from 1.55 to 2.56~\cite{kim2025development}. These results supported consistency in the large-scale spatial patterns and overall scale of the calculated metrics across independently constructed populations. Building on this initial implementation, the present study establishes a reproducible analysis workflow, quantifies the dependence of the resulting risk metrics and computational cost on spatial resolution, and demonstrates its application to historical and future-scenario analyses.

\subsection{Intended Applications and Users}

As presented in Fig.~\ref{fig:integrated_context}, within Korea's integrated collision-risk analysis framework, KASI is expected to provide SSA-related services under the coordination of the Korea AeroSpace Administration (KASA), supporting decisions across multiple levels, from satellite operations to national policy. The intended users include KASA and KASI themselves, other public institutions operating space assets, such as the Korea Aerospace Research Institute (KARI), and commercial satellite operators.

For the 3D-cell model specifically, the primary intended users are national SSA and space-environment analysts, as well as policy and planning teams. Supported tasks include monitoring the current catalog-conditioned environment, comparing historical snapshots and candidate deployment scenarios, identifying spatial regions with concentrated object populations or collision flux, communicating environmental trends, and prioritizing cases for higher-fidelity follow-up analysis. For example, within the broader Korean SSA framework, KASI is conducting ongoing research on space-object catalog construction~\cite{kasi_risk_catalog}. As domestic catalog products become available, they can serve as upstream inputs to the 3D-cell workflow. This would enable routinely updated products, such as daily maps of regional space traffic represented by spatial density, together with collision-flux indicators for selected orbital regions or national space assets. Such products could support continuous environmental monitoring and provide a common quantitative basis for operational, planning, and policy-level assessments.

A secondary user group includes satellite operators, mission designers, and researchers, who may incorporate the model into their own workflows. For example, when a satellite operator is considering multiple candidate orbits for a new satellite deployment, the 3D-cell model can be used to compare the collision exposure associated with each candidate orbit under a specified projected future environment.}

\section{\rev{3D-Cell Analysis Method}}\label{sec:method}

\rev{This section introduces the mathematical formulations of the 3D-cell analysis method used throughout this study. It first describes the analysis workflow and the boundary between external inputs and the analysis modules. It then specifies the coordinate convention and cell geometry and derives the cell-passage residence, spatial density, impact flux, and collision probability.

\subsection{Analysis Workflow}

Figure \ref{fig:framework} separates the externally supplied population and target inputs from the functional modules of the 3D-cell analysis. The gray block represents the external inputs, whereas the blue blocks denote the functional analysis modules. Following the concept of the previous research~\cite{liou2004cube,klinkrad2006space,esa2020master_manual}, the model partitions space into cells as spatial analysis units and assumes that interactions occur only between objects occupying the same cell.

The ``Cell Definition'' module partitions the analysis domain according to the selected spatial bounds and cell widths. The ``Cell Passage Event \& Spatial Density'' module determines each object's boundary-crossing times and cell residence durations and then converts the accumulated residence probabilities into spatial density. The ``Impact Flux \& Collision Probability'' module combines this density field with the target residence information to calculate impact flux and collision probability. Post-analysis products support visualization and refinement of the analysis configuration. If the grid is too coarse to resolve localized variations in spatial density or target-specific metrics, the cell widths can be refined and the calculation repeated.

\subsection{Coordinate Convention and Cell Definition}

This subsection describes the ``Cell Definition'' module. The cell formulation can be applied in any consistently defined Earth-centered inertial (ECI) reference frame, such as TEME or GCRF. In the selected reference frame, $r=\sqrt{x^2+y^2+z^2}$ is the geocentric radius, $\delta=\arcsin(z/r)$ is the geocentric declination, and $\alpha=\operatorname{atan2}(y,x)$ is the right ascension.

The user defines the analysis bounds through the intervals $(r_{\min}, r_{\max})$, $(\delta_{\min}, \delta_{\max})$, and $(\alpha_{\min}, \alpha_{\max})$, together with the cell widths $(\Delta r, \Delta \delta, \Delta \alpha)$, as illustrated in Fig.~\ref{fig:cell_definition}. These choices determine the domain coverage, spatial resolution, and computational cost. Wider domains and smaller cells increase the computational burden, whereas insufficient coverage or overly coarse cells can obscure relevant spatial structures. The bounds and cell widths should therefore be selected according to the target scenario and evaluated through resolution analysis.

\begin{figure}[hbt!]
    \centering
    \includegraphics[page=3,width=0.92\linewidth]{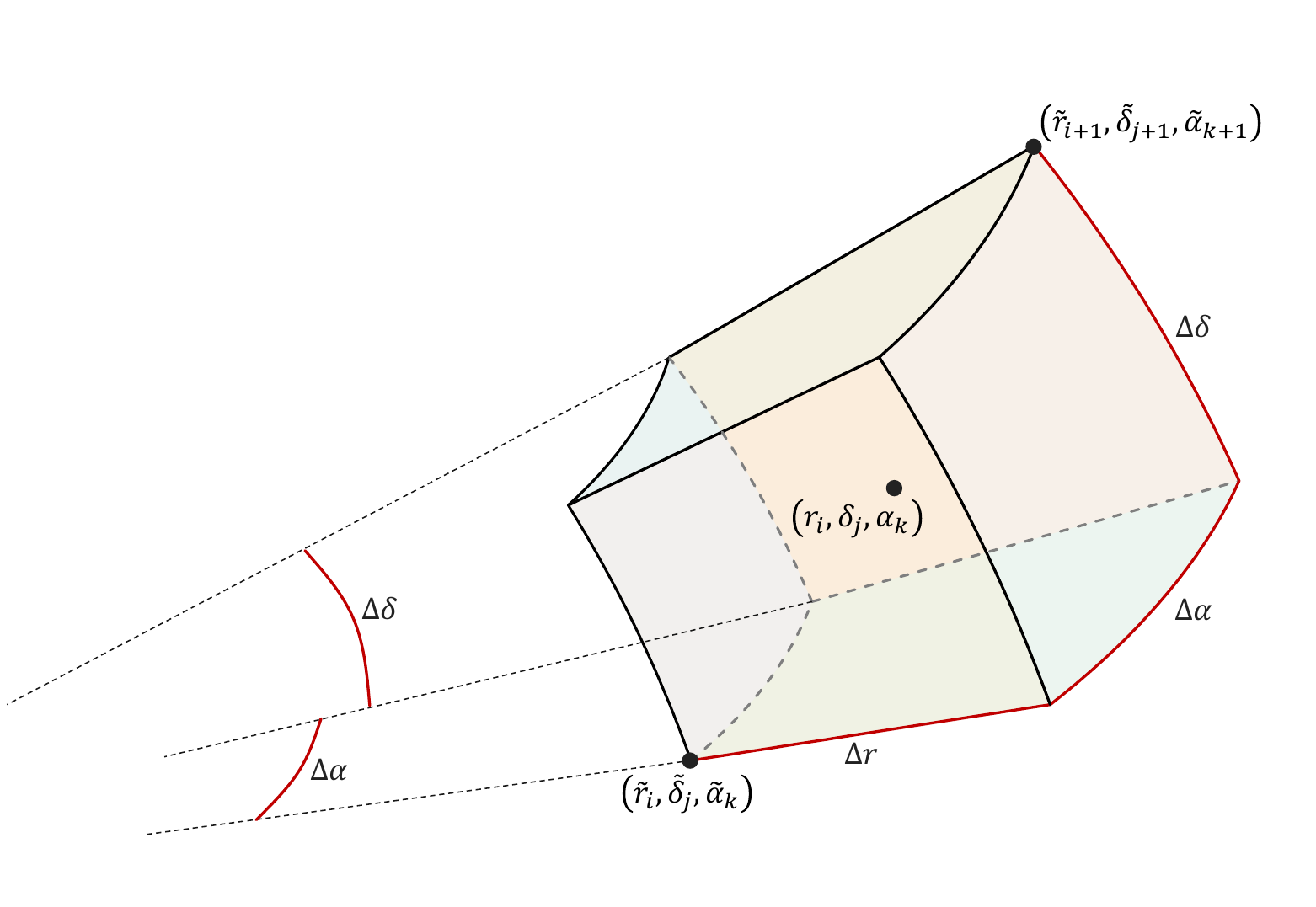}
    \caption{\rev{Spherical 3D-cell geometry in radial, declination, and right-ascension coordinates $(r,\delta,\alpha)$. The angular widths $\Delta\delta$ and $\Delta\alpha$ are illustrated schematically and do not represent arc lengths, whereas $\Delta r$ denotes the radial cell width. Dashed lines indicate occluded cell boundaries, and dotted lines indicate angular directions.}}
    \label{fig:cell_definition}
\end{figure}

Based on these inputs, the cell boundaries are defined as follows~\cite{klinkrad2006space}:}
\begin{equation}
    \tilde{r}_i=r_{\min}+i\Delta r,\quad i=0,1,\dots,I,
\end{equation}
\begin{equation}
    \tilde{\delta}_j=\delta_{\min}+j\Delta \delta,\quad j=0,1,\dots,J,
\end{equation}
\begin{equation}
    \tilde{\alpha}_k=\alpha_{\min}+k\Delta \alpha,\quad k=0,1,\dots,K,
\end{equation}
where $I$, $J$, and $K$ are defined as
\begin{equation}
    I=\left\lfloor \frac{r_{\max}-r_{\min}}{\Delta r}\right\rfloor,\quad
    J=\left\lfloor \frac{\delta_{\max}-\delta_{\min}}{\Delta \delta}\right\rfloor,\quad
    K=\left\lfloor \frac{\alpha_{\max}-\alpha_{\min}}{\Delta \alpha}\right\rfloor.
\end{equation}
We assign an index triple $(i,j,k)$ to each cell, where $i=0,\dots,I-1$, $j=0,\dots,J-1$, and $k=0,\dots,K-1$. Each cell $(i,j,k)$ is bounded by the surfaces $\tilde{r}_i$ and $\tilde{r}_{i+1}$, $\tilde{\delta}_j$ and $\tilde{\delta}_{j+1}$, and $\tilde{\alpha}_k$ and $\tilde{\alpha}_{k+1}$, resulting in a total of $I\times J\times K$ cells. \rev{The position vector of the center of cell $(i,j,k)$ is}
\begin{equation}
    \mathbf{r}_{ijk}=r_i\left[\cos\delta_j\cos\alpha_k, \cos\delta_j\sin\alpha_k,\sin\delta_j\right]^\top,
\end{equation}
where
\begin{equation}
    r_i=r_{\min}+\left(i+\frac{1}{2}\right)\Delta r,\quad i=0,1,\dots,I-1,
\end{equation}
\begin{equation}
    \delta_j=\delta_{\min}+\left(j+\frac{1}{2}\right)\Delta \delta,\quad j=0,1,\dots,J-1,
\end{equation}
\begin{equation}
    \alpha_k=\alpha_{\min}+\left(k+\frac{1}{2}\right)\Delta \alpha,\quad k=0,1,\dots,K-1.
\end{equation}

\rev{The volume of cell $(i,j,k)$ is calculated as
\begin{equation}\label{eq:cell_volume}
V_{ijk}=\frac{2}{3}\left(3r_i^2+\frac{\Delta r^2}{4}\right)
\cos\delta_j\sin\left(\frac{\Delta \delta}{2}\right)\Delta\alpha\Delta r,
\end{equation}
where $\Delta\delta$ and $\Delta\alpha$ are expressed in radians. The bracketed radial factor has units of km$^2$, and multiplication by $\Delta r$ gives the cell volume in km$^3$.}

\subsection{\rev{Cell-Passage Residence and Collision Metric}}

\rev{This subsection describes the ``Cell Passage Event \& Spatial Density'' and ``Impact Flux \& Collision Probability'' modules.} A \rev{CPE} is defined as an event in which an object crosses a cell boundary. \rev{The resulting CPEs identify the times at which the object crosses the cell boundaries and thereby determine the duration for which the object resides in each cell over one orbital period. For an object under Keplerian dynamics, the true anomalies corresponding to crossings of the boundaries of constant radius $\tilde{r}_i$, constant declination $\tilde{\delta}_j$, and constant right ascension $\tilde{\alpha}_k$ are given by}
\begin{equation}
    f_{\tilde{r}_i,1} = \arccos\left(\frac{a(1-e^2)-\tilde{r}_i}{e\tilde{r}_i}\right),\quad f_{\tilde{r}_i,2}=2\pi-f_{\tilde{r}_i,1},\quad i\in\mathcal{I}_{\mathrm{CPE}},
\end{equation}
\begin{equation}
    f_{\tilde{\delta}_j,1}=\arcsin\left(\frac{\sin\tilde{\delta}_j}{\sin\mathrm{inc}}\right)-\omega,\quad f_{\tilde{\delta}_j,2}=\pi-\arcsin\left(\frac{\sin\tilde{\delta}_j}{\sin\mathrm{inc}}\right)-\omega,\quad j\in\mathcal{J}_{\mathrm{CPE}},
\end{equation}
\begin{equation}
    f_{\tilde{\alpha}_k}=
    \begin{cases}
        \arctan\left(\frac{\tan(\tilde{\alpha}_k-\Omega)}{\cos\mathrm{inc}}\right)-\omega, & \text{if } \cos(\tilde{\alpha}_k-\Omega)\geq 0, \\
        \arctan\left(\frac{\tan(\tilde{\alpha}_k-\Omega)}{\cos\mathrm{inc}}\right)-\omega+\pi, & \text{if } \cos(\tilde{\alpha}_k-\Omega)<0,
    \end{cases}\rev{\pmod{2\pi}},\quad k=0,1,\dots,K,
\end{equation}
where
\begin{equation}
    \mathcal{I}_{\mathrm{CPE}}=\left\{i=0,1,\dots,I:a(1-e)<\tilde{r}_i<a(1+e)\right\},
\end{equation}
\begin{equation}
    \mathcal{J}_{\mathrm{CPE}}=\left\{j=0,1,\dots,J:|\sin\tilde{\delta}_j|<|\sin\mathrm{inc}|\right\},
\end{equation}
and $a$, $e$, $\mathrm{inc}$, $\omega$, and $\Omega$ are the classical orbital elements of the object: semi-major axis, eccentricity, inclination, argument of periapsis, and right ascension of the ascending node, respectively. \rev{The index sets $\mathcal{I}_{\mathrm{CPE}}$ and $\mathcal{J}_{\mathrm{CPE}}$ restrict the radial and declination boundaries to those crossed by the orbit.

The eccentric anomaly, $E$, associated with true anomaly $f$ is
\begin{equation}
E(f)=2\operatorname{atan2}\left(\sqrt{1-e}\sin\frac{f}{2},\sqrt{1+e}\cos\frac{f}{2}\right).
\end{equation}
The elapsed time from periapsis to a specified true anomaly is then
\begin{equation}
t(f)=\sqrt{\frac{a^3}{\mu}}\left[E(f)-e\sin E(f)\right]\pmod{T},\qquad T=2\pi\sqrt{\frac{a^3}{\mu}},
\end{equation}
where $\mu$ is the gravitational parameter of the central body, here the Earth, and $T$ is the orbital period. Sorting the combined radial and angular boundary-crossing times partitions one orbital period into cell-residence intervals. For object $l$, the total duration of the intervals assigned to cell $(i,j,k)$ is denoted by $t_{ijkl}$. The residence probability of object $l$ in cell $(i,j,k)$, i.e., the probability of finding object $l$ in that cell at a randomly sampled time, is then}
\begin{equation}\label{eq:resident prob}
p_{ijkl}=\frac{t_{ijkl}}{T_l},\quad T_l=2\pi\sqrt{\frac{a_l^3}{\mu}},\quad l=1,2,\dots,L,
\end{equation}
where $T_l$ and $a_l$ are the orbital period and semi-major axis of object $l$, respectively, and $L$ is the total number of objects.

\rev{The aggregate spatial density in cell $(i,j,k)$ formed by the $L$ objects is}
\begin{equation}
D_{ijk}=\frac{1}{V_{ijk}}\sum_{l=1}^{L}p_{ijkl}.
\end{equation}
\rev{The same procedure gives the residence probability $p^{(q)}_{ijk}$ of target $q$ in cell $(i,j,k)$. Hereafter, ``risk objects'' refer to the population that poses collision risk to the target, whereas ``targets'' denote the objects of interest for which collision exposure is evaluated. For example, when assessing a Korean space asset, the catalogued objects constitute the risk-object population, while the Korean asset is treated as the target.

For target $q$, the velocity-independent target-weighted density overlap is defined as
\begin{equation}
S_q=\sum_{i=0}^{I-1}\sum_{j=0}^{J-1}\sum_{k=0}^{K-1}p^{(q)}_{ijk}D_{ijk}.
\label{eq:overlap}
\end{equation}
This quantity samples the catalog-conditioned density field according to the target's time-averaged cell residence and isolates the spatial component of collision exposure. At the present level of model reduction, a prescribed effective relative speed $v_c$ converts this overlap into collision flux:\footnote{\rev{The calculations reported in this study use $v_c=10$ km/s as a macroscopic reference value rather than a target-specific encounter-speed prediction. If an application instead specifies an effective-speed interval $[v_L,v_U]$, the linear relation gives $F_q\in[S_qv_L,S_qv_U]$.}}
\begin{equation}
F_q=S_qv_c.
\label{eq:flux}
\end{equation}

Following the kinetic theory of gases~\cite{klinkrad2006space}, the expected number of impacts on target $q$ over an interval $\Delta t$, for collision cross-section $A_c$, is
\begin{equation}\label{eq:mean_impact_count}
c_q=F_qA_c\Delta t.
\end{equation}
Let $N_q$ denote the number of impacts on target $q$ during $\Delta t$. Under a Poisson model with mean $c_q$, the probability of at least one impact is the complement of the zero-impact probability:
\begin{equation}\label{eq:risk}
P(N_q\geq1)=1-P(N_q=0)=1-\exp(-c_q).
\end{equation}
For $c_q\ll1$, this probability is approximately $c_q$. The exact Poisson expression is used throughout this study, while $c_q$ is reported separately because expected impact counts are additive across disjoint catalog groups.}

\rev{
\section{Cell-Resolution Sensitivity and Computational Performance}\label{sec:numerical_study}

This section evaluates how spatial cell resolution affects the outputs of the 3D-cell analysis and reports the associated computational performance. It first defines the catalog inputs, target set, computational environment, and cell-resolution design, and then reports numerical consistency checks, collision-metric sensitivity, and runtime and memory measurements.

\subsection{Data and Experimental Setup}\label{sec:numerical_study_1}

For each year from 2016 through 2025, all Space-Track GP records~\cite{spacetrack} with epochs in the half-open August 1 to September 1 window were collected to construct a fixed September 1 annual snapshot. For each NORAD catalog identifier, the TLE nearest September 1 00:00:00 UTC was selected and propagated once to that common epoch using SDP4 or SGP4, as appropriate, through the SGP4 Python library~\cite{rhodes2025sgp4}. Across the ten snapshots, the annual median TLE-to-snapshot propagation ranges from 0.22 to 0.58 d, while the annual 95th percentile ranges from 2.90 to 5.41 d; a sparse tail of stale records approaches the 31-d window limit. This section uses the 2025 snapshot for the resolution and performance analyses, whereas Section~\ref{sec:case_study_1} uses all ten annual snapshots for the historical analysis.

The population classification follows an ordered, exhaustive rule. Object-name prefixes first identify Starlink and OneWeb objects, as well as the FENGYUN 1C, COSMOS 2251, IRIDIUM 33, and COSMOS 1408 debris families.\footnote{\rev{The named-debris category is an analysis-specific grouping rather than a native Space-Track object class. Name prefixes are used to isolate debris associated with documented collision and anti-satellite events for attribution~\cite{iridium_cosmos_collision,chinese_asat,russian_asat}.}} Remaining records are assigned according to the Space-Track object type as other payload, rocket body, or other debris, while records with unmatched, blank, or TBA object types are assigned to the unknown/TBA group. Each accepted object is therefore assigned to exactly one group. Figure~\ref{fig:density_population} pairs the primary 2025 spatial-density field expressed in terms of geocentric altitude $r-R_\mathrm{E}$ with population counts binned by semi-major axis relative to the reference Earth radius, $a-R_\mathrm{E}$, where $R_\mathrm{E}=6378.137$ km is the reference Earth radius used in the calculations. The strongest lower-LEO density bands coincide with narrow Starlink semi-major-axis peaks, while the density feature near 1,200 km aligns with OneWeb. Broader payload, debris, rocket-body, and unknown/TBA populations form a more diffuse background.

\begin{figure}[hbt!]
    \centering
    \subfloat[Spatial density.]{\includegraphics[page=4,width=0.96\linewidth]{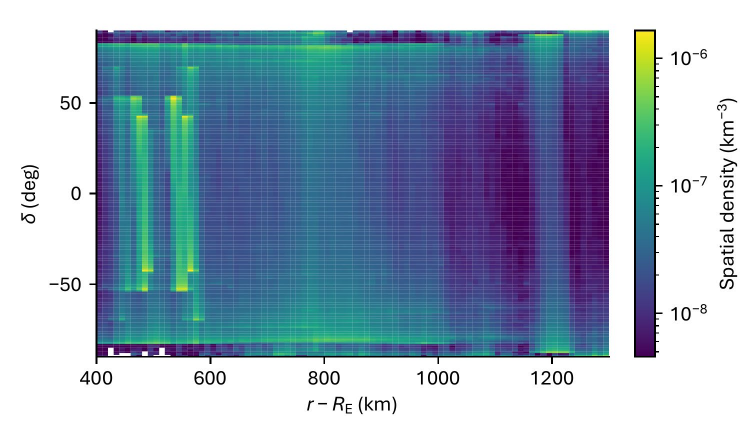}\label{fig:density_2025}}\\[0.4em]
    \subfloat[Population by semi-major axis relative to the reference Earth radius.]{\includegraphics[page=5,width=0.96\linewidth]{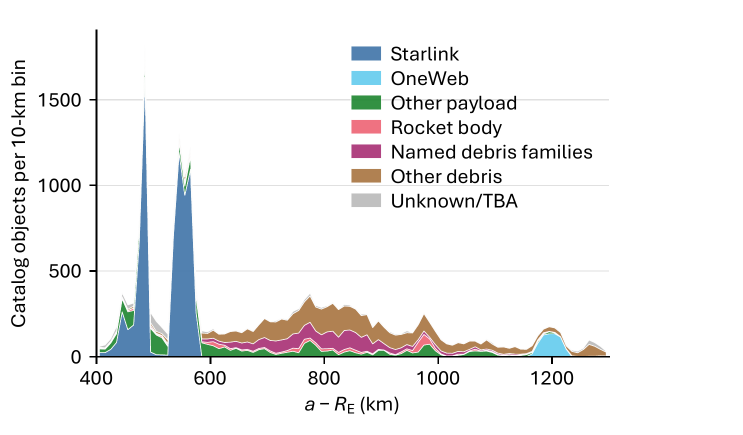}\label{fig:population_altitude_2025}}
    \caption{Spatial density and population distribution of the 2025 catalog snapshot.}
    \label{fig:density_population}
\end{figure}

Table~\ref{tab:targets} summarizes six targets with complementary analytical roles. The first four are synthetically defined orbits. Circular 1 samples a dense lower-LEO constellation region, Circular 2 represents a 700-km sun-synchronous orbit (SSO), and Circular 3 samples a lower-inclination upper-LEO region. The eccentric LEO (ELEO) target is a synthetic critical-inclination orbit with perigee and apogee altitudes of approximately 450 and 1,250 km, respectively, designed to span a broad range of LEO altitudes. The remaining two targets were selected both for their relevance as Korean space assets and to complement the synthetic targets with distinct altitude--inclination regimes. KOMPSAT-3A (NORAD 40536) is a Korean Earth-observation satellite operating in an SSO~\cite{kari_kompsat3a} and provides a low-altitude near-polar case. DOORY-SAT (NORAD 58500) is associated with Hanwha Systems' Korean small-SAR Earth-observation mission~\cite{hanwha2024_smallsar,spacetrack} and provides a mid-altitude case at approximately $47^\circ$ inclination. The orbital elements of the two catalogued targets are obtained from the corresponding Space-Track records at the 2025 analysis epoch~\cite{spacetrack}. Each catalogued asset is removed from the risk-object population before evaluating its own exposure. All orbital elements are expressed with respect to the TEME frame used in this study; $a$, $e$, $\mathrm{inc}$, $\omega$, and $\Omega$ denote semi-major axis, eccentricity, inclination, argument of periapsis, and right ascension of the ascending node, respectively.

\begin{table}[hbt!]
\caption{\rev{Orbital elements of the six analysis targets.}}\label{tab:targets}
\rev{
\centering

{\setlength{\tabcolsep}{4pt}\renewcommand{\arraystretch}{1.12}%
\begin{tabular}{@{}>{\raggedright\arraybackslash}m{2.2cm}>{\centering\arraybackslash}m{1.45cm}>{\centering\arraybackslash}m{1.75cm}>{\centering\arraybackslash}m{1.25cm}>{\centering\arraybackslash}m{1.25cm}>{\centering\arraybackslash}m{1.25cm}@{}}
\toprule
\textbf{Target} & $a$ (km) & $e$ & $\mathrm{inc}$ ($^\circ$) & $\omega$ ($^\circ$) & $\Omega$ ($^\circ$) \\
\midrule
Circular 1 & 6,878.14 & $1.0\times10^{-4}$ & 60.00 & 0.00 & 0.00 \\
Circular 2 & 7,078.14 & $1.0\times10^{-4}$ & 98.19 & 0.00 & 0.00 \\
Circular 3 & 7,378.14 & $1.0\times10^{-4}$ & 30.00 & 0.00 & 0.00 \\
ELEO & 7,228.14 & $5.534\times10^{-2}$ & 63.40 & 0.00 & 0.00 \\
\addlinespace[2pt]
KOMPSAT-3A & 6,819.43 & $8.684\times10^{-4}$ & 97.67 & 45.25 & 213.17 \\
DOORY-SAT & 7,014.96 & $1.925\times10^{-3}$ & 47.00 & 72.51 & 269.52 \\
\botrule
\end{tabular}
}
}
\end{table}

Table~\ref{tab:computing_environment} summarizes the hardware and software environment used for the computational performance measurements.

\begin{table}[hbt!]
\caption{\rev{Computational environment for the numerical experiments.}}\label{tab:computing_environment}
\rev{
\centering
{\setlength{\tabcolsep}{6pt}\renewcommand{\arraystretch}{1.10}%
\begin{tabular}{@{}>{\raggedright\arraybackslash}m{3.1cm}>{\raggedright\arraybackslash}m{7.4cm}@{}}
\toprule
\textbf{Item} & \textbf{Setting} \\
\midrule
Processor & Apple M4 Pro, 14 CPU cores \\
Memory & 64 GB \\
Language & Python 3.13.9 \\
Packages & NumPy 2.3.5; Numba 0.62.1 \\
\botrule
\end{tabular}
}
}
\end{table}

\subsection{Parameter Settings for Resolution Analysis}\label{sec:numerical_study_2}

The numerical study covers geocentric altitudes from 400 to 1,300 km, declinations from $-90^\circ$ to $90^\circ$, and right ascensions from $-180^\circ$ to $180^\circ$. The reference and tested parameter values are summarized in Table~\ref{tab:parameter_ranges}. The reference cell widths are $(\Delta r,\Delta\delta,\Delta\alpha)=(10\ \mathrm{km},1^\circ,360^\circ)$, selected largely based on the baseline settings used in MASTER, with the right-ascension dimension aggregated into a single $360^\circ$ ring~\cite{klinkrad2006space}. For controlled comparison across spatial resolutions, the effective relative speed, collision cross-section, and analysis interval are fixed at $v_c=10$ km/s, $A_c=1$ m$^2$, and $\Delta t=365$ d, respectively.

The sensitivity design includes eight widths for each coordinate: $\Delta r=45$--0.25 km, $\Delta\delta=4^\circ$--$0.01^\circ$, and $\Delta\alpha=360^\circ$--$1^\circ$, with the exact discrete values listed in Table~\ref{tab:parameter_ranges}. Every tested width exactly tiles its corresponding 900-km, $180^\circ$, or $360^\circ$ domain span. Each configuration varies only one of $\Delta r$, $\Delta\delta$, or $\Delta\alpha$ from the reference setting, allowing the influence of each cell-width parameter to be examined independently.

\begin{table}[hbt!]
\caption{\rev{Reference settings and tested spatial-resolution values.}}\label{tab:parameter_ranges}
\rev{
\centering
\small

{\setlength{\tabcolsep}{3.5pt}\renewcommand{\arraystretch}{1.14}%
\begin{tabular}{@{}>{\centering\arraybackslash}m{2.2cm}>{\centering\arraybackslash}m{1.8cm}>{\centering\arraybackslash}m{7.4cm}@{}}
\toprule
\textbf{Parameter} & \textbf{Reference} & \textbf{Tested values} \\
\midrule
$\Delta r$ (km) & 10 & 45, 20, 10, 5, 2.5, 1, 0.5, 0.25 \\
$\Delta\delta$ ($^\circ$) & 1 & 4, 2, 1, 0.5, 0.25, 0.1, 0.05, 0.01 \\
$\Delta\alpha$ ($^\circ$) & 360 & 360, 180, 90, 30, 10, 5, 2.5, 1 \\
$v_c$ & 10 km/s & Fixed \\
$A_c$, $\Delta t$ & 1 m$^2$, 365 d & Fixed \\
\botrule
\end{tabular}
}
}
\end{table}

\subsection{Results}\label{sec:numerical_study_3}

\subsubsection{Collision Metrics and Resolution Sensitivity}

Figure~\ref{fig:resolution_sensitivity} presents the sensitivity results for variations in all three cell-width parameters using the 2025 catalogue. Each response is normalized by the corresponding value for that target at the reference setting. The horizontal axes show the tested cell widths on linear scales, while the normalized responses are shown on logarithmic vertical scales. Lines connect the discrete cases to indicate the numerical trends, and dashed vertical lines mark the reference widths.

\begin{figure}[hbt!]
    \centering
    \subfloat[Radial width.]{\includegraphics[page=6,width=0.92\linewidth]{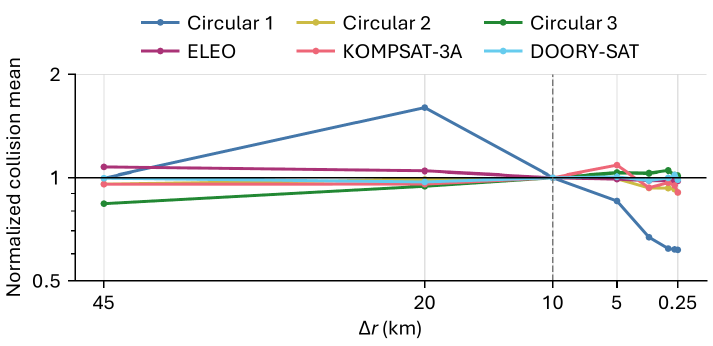}\label{fig:resolution_radial}}\\[0.25em]
    \subfloat[Declination width.]{\includegraphics[page=7,width=0.92\linewidth]{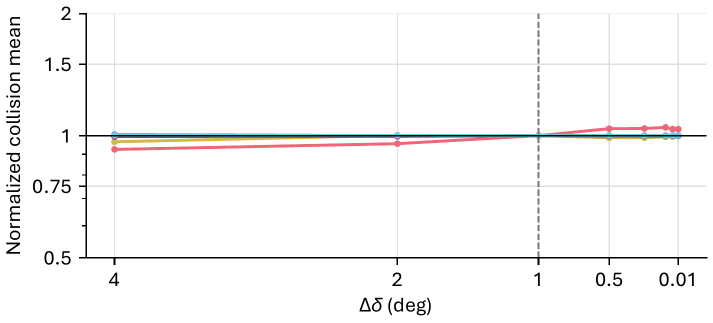}\label{fig:resolution_declination}}\\[0.25em]
    \subfloat[Right-ascension width.]{\includegraphics[page=8,width=0.7\linewidth]{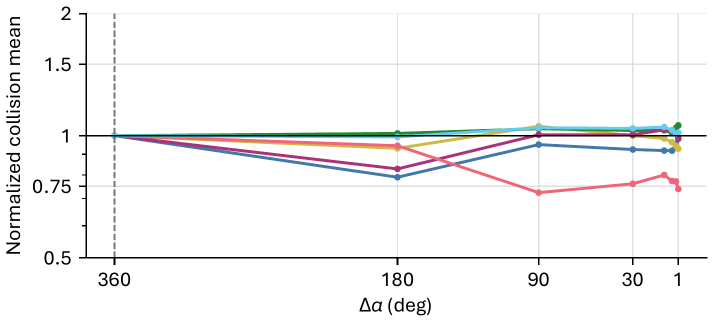}\label{fig:resolution_ra}}
    \caption{\rev{Normalized expected impact-count responses to spatial-resolution variations for the 2025 catalog.}}
    \label{fig:resolution_sensitivity}
\end{figure}

Across the six targets, the normalized radial, declination, and right-ascension responses range from 0.615--1.599, 0.925--1.048, and 0.724--1.060 relative to the reference values, respectively. The results indicate that sensitivity to spatial resolution is both coordinate- and target-dependent, with the largest variations observed for radial and right-ascension refinement. The response curves are not uniformly monotonic, indicating that reducing the cell width does not necessarily produce a monotonic change in the target metric. This behavior can arise because changes in a global cell width shift cell boundaries relative to localized orbital structures.

Table~\ref{tab:resolution_summary} summarizes the change in the expected impact count ($c_q$ in Eq.~\eqref{eq:mean_impact_count}) between the reference setting and the finest tested width for each target--coordinate combination. It also reports the maximum absolute value of the last two successive relative changes, evaluated over the three finest tested widths for each coordinate, denoted as the Max. final-step change. These local changes are all below 5\%, ranging from less than 0.01\% to 4.54\%. The largest reference-to-finest change occurs for the radial refinement of Circular 1, for which the finest-grid result is 38.45\% lower than the reference result. The right-ascension refinement of KOMPSAT-3A also produces a substantial decrease of 26.11\%, whereas refinement in declination produces comparatively small changes for most targets.

\begin{table}[hbt!]
\caption{\rev{Finest-width changes and local final-step variations for the 2025 catalog.}}\label{tab:resolution_summary}
\rev{
\centering

{\setlength{\tabcolsep}{3.2pt}\renewcommand{\arraystretch}{1.05}%
\begin{tabular}{@{}>{\raggedright\arraybackslash}m{2.6cm}>{\centering\arraybackslash}m{1.4cm}>{\centering\arraybackslash}m{3.0cm}>{\centering\arraybackslash}m{3.6cm}@{}}
\toprule
\textbf{Target} & \textbf{Axis} & \textbf{Finest vs. reference (\%)} & \textbf{Max. final-step change (\%)} \\
\midrule
\multirow{3}{*}{Circular 1}
& $\Delta r$ & $-38.45$ & 0.58 \\
& $\Delta\delta$ & $+0.07$ & $<0.01$ \\
& $\Delta\alpha$ & $-1.21$ & 3.87 \\
\addlinespace[2pt]

\multirow{3}{*}{Circular 2}
& $\Delta r$ & $-9.41$ & 2.20 \\
& $\Delta\delta$ & $-0.11$ & 0.41 \\
& $\Delta\alpha$ & $-7.20$ & 2.01 \\
\addlinespace[2pt]

\multirow{3}{*}{Circular 3}
& $\Delta r$ & $+1.23$ & 2.97 \\
& $\Delta\delta$ & $-0.10$ & 0.03 \\
& $\Delta\alpha$ & $+6.01$ & 2.87 \\
\addlinespace[2pt]

\multirow{3}{*}{ELEO}
& $\Delta r$ & $-2.08$ & 0.07 \\
& $\Delta\delta$ & $+0.03$ & $<0.01$ \\
& $\Delta\alpha$ & $-1.78$ & 3.44 \\
\addlinespace[2pt]

\multirow{3}{*}{KOMPSAT-3A}
& $\Delta r$ & $-9.36$ & 4.54 \\
& $\Delta\delta$ & $+3.74$ & 0.94 \\
& $\Delta\alpha$ & $-26.11$ & 4.14 \\
\addlinespace[2pt]

\multirow{3}{*}{DOORY-SAT}
& $\Delta r$ & $-1.30$ & 2.81 \\
& $\Delta\delta$ & $+0.09$ & $<0.01$ \\
& $\Delta\alpha$ & $+1.81$ & 2.12 \\
\botrule
\end{tabular}
}
}
\end{table}

\subsubsection{Computational Performance}

Computational performance was recorded during the same 22 one-axis 2025 resolution calculations used in this section. Figure~\ref{fig:performance_config_time} separates the refinement cases into two branches: $\Delta r$ and $\Delta\delta$ on one branch, and $\Delta\alpha$ on the other. This distinction reflects the fact that configurations with the same total number of cells can exhibit substantially different runtimes. At 64,800 cells, for example, the radial, declination, and right-ascension cases required 7.03, 9.30, and 1.74 s, respectively. Radial and declination refinements involve substantially more one-dimensional boundary tests and orbit-passage intervals than right-ascension refinement at the same total cell count. In contrast, the peak memory usage shown in Fig.~\ref{fig:performance_config_memory} follows the allocated grid size more directly.

\begin{figure}[hbt!]
    \centering
    \subfloat[Cell count vs. computing time.]{\includegraphics[page=9,width=0.48\linewidth]{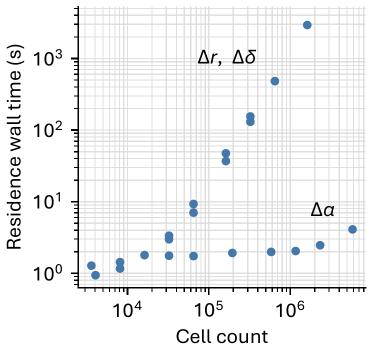}\label{fig:performance_config_time}}\hfill
    \subfloat[Cell count vs. peak memory.]{\includegraphics[page=10,width=0.48\linewidth]{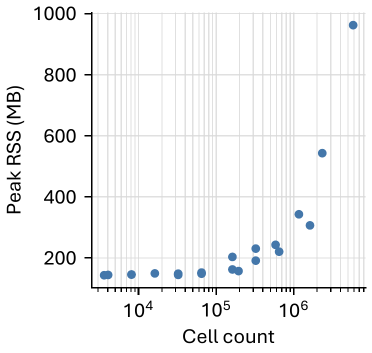}\label{fig:performance_config_memory}}
    \caption{Computational performance of the residence calculation across the tested resolution configurations.}
    \label{fig:performance}
\end{figure}

\subsection{Discussion}\label{sec:numerical_discussion}

The resolution study shows that the 3D-cell outputs can exhibit non-negligible sensitivity to spatial discretization. Across all tested widths, the normalized expected impact counts range from 0.615 to 1.599 of their reference values, while the changes at the finest tested widths range from $-38.5\%$ to $+6.01\%$. Spatial resolution can therefore materially affect quantitative estimates for some targets and coordinates. Nevertheless, the variations observed in the present study remain within the same order of magnitude. For example, if a reference calculation yields an expected impact count of $c_q=2$ over a specified analysis interval, the full normalized range observed in this study corresponds to approximately 1.23--3.20 expected impacts.

These results are consistent with the intended use of the 3D-cell workflow as a screening-level comparative tool. The reference setting $(\Delta r,\Delta\delta,\Delta\alpha)=(10\ \mathrm{km},1^\circ,360^\circ)$ provides a computationally efficient basis for comparative analyses, but resolution sensitivity should be considered when interpreting the resulting metrics. Target-specific applications requiring more precise quantitative estimates should employ further grid refinement and explicitly characterize discretization sensitivity. An adaptive or locally refined grid could reduce the computational cost of obtaining higher spatial resolution where needed and is a priority for future development.

The effective-speed assumption can be treated separately from spatial-resolution dependence because $F_q=S_qv_c$. Once the velocity-independent spatial overlap $S_q$ has been calculated, an application-specific effective relative speed can rescale the resulting flux without repeating the spatial calculation. For example, increasing $v_c$ from 10 to 20 km/s doubles both $F_q$ and the expected impact count $c_q$, while the corresponding collision probability follows the Poisson relation in Eq.~\eqref{eq:risk}. Thus, an application may adopt a different or deliberately conservative effective-speed assumption according to its intended use while retaining the same spatial-overlap calculation.

The right-ascension results require additional consideration because they are inherently dependent on both the target and the catalog snapshot. The geocentric right ascension of an object's instantaneous position evolves over each orbit, while $J_2$-driven nodal precession changes the orientation of the orbital plane over longer time scales. The present fixed-epoch residence calculation represents orbital motion within a fixed orbital-plane orientation but does not model the long-horizon evolution of that orientation. The fine-$\Delta\alpha$ responses therefore characterize snapshot-conditioned right-ascension structure rather than universal resolution behavior, and their nonmonotonic variation shows that decreasing $\Delta\alpha$ alone does not necessarily produce a stable macroscopic result. Setting $\Delta\alpha=360^\circ$ provides a ring-aggregated representation suitable for macroscopic screening when local right-ascension structure is not of interest.

The computational results further illustrate the trade-off between spatial resolution and analysis cost. Fine radial and declination grids substantially increase residence-calculation time, whereas right-ascension refinement primarily increases memory demand. For analyses involving only a small number of catalog snapshots or scenarios, finer spatial resolutions may therefore be computationally practical. In contrast, analyses involving many catalog snapshots, deployment scenarios, or repeated environment updates require the residence calculation to be repeated for each supplied population, and the cumulative computational burden can become significant. Spatial resolution should therefore be selected by balancing the required level of numerical detail against the number of environments or scenarios to be evaluated.

\section{Case Studies}\label{sec:case_study}

This section examines historical and prospective changes in the collision metrics for the six targets defined in Section~\ref{sec:numerical_study_1}. The historical analysis compares ten annual catalog snapshots from 2016 to 2025, whereas the prospective case study augments the 2025 catalog environment with the population proposed for SpaceX's ODC~\cite{spacex2026odc_response}.

\subsection{Annual Snapshots: 2016--2025}\label{sec:case_study_1}

To investigate changes in the LEO environment from 2016 to 2025, we analyze the ten September 1 catalog snapshots constructed using the procedure described in Section~\ref{sec:numerical_study_1}. Across these annual snapshots, the catalog population increases from 15,723 objects in 2016 to 28,540 in 2025, as shown in Fig.~\ref{fig:population_history}.

\begin{figure}[hbt!]
    \centering
    \includegraphics[page=11,width=0.98\linewidth]{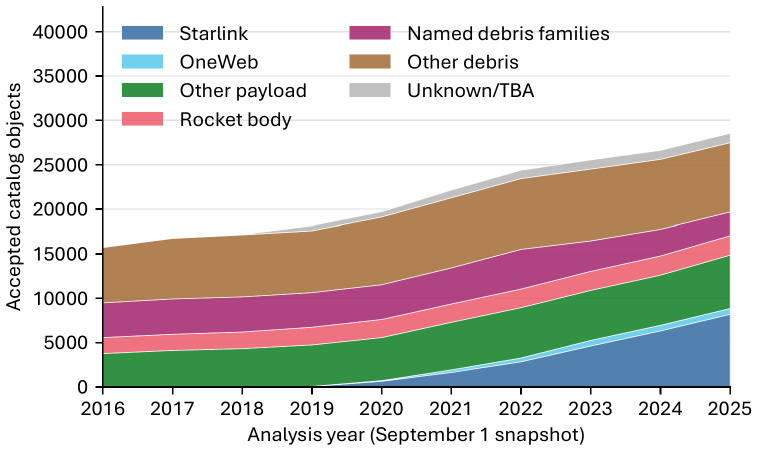}
    \caption{Annual catalog population from 2016 to 2025.}
    \label{fig:population_history}
\end{figure}

The signed density difference in Fig.~\ref{fig:density_difference} shows strong positive (red) bands near the lower-LEO constellation shells and around 1,200 km, whereas negative (blue) regions partly reflect catalog turnover and the redistribution or decay of previously tracked populations. A group-resolved analysis attributes the prominent negative band near 830--870 km primarily to FENGYUN 1C debris. The 2007 anti-satellite breakup occurred in an 845--865-km, $98.6^\circ$ orbit, and a NASA review found that the resulting debris cloud still accounted for approximately 36\% of the catalogued spatial density at 840--860 km in June 2023~\cite{anzmeador2023debrisclouds}. In the present snapshots, the number of accepted FENGYUN 1C debris records decreases from 2,550 in 2016 to 1,912 in 2025, while the number within the 830--870-km semi-major-axis-based altitude range decreases from 561 to 385. Of the 722 identifiers accepted in 2016 but absent in 2025, 419 have SATCAT decay dates no later than the 2025 analysis epoch. Among the objects within the 2016 band that are retained in both snapshots, 274 have shifted below 830 km by 2025. The negative band reflects a combination of documented decay, downward orbital redistribution, and snapshot-specific catalog turnover.

\begin{figure}[hbt!]
    \centering
    \includegraphics[page=12,width=0.97\linewidth]{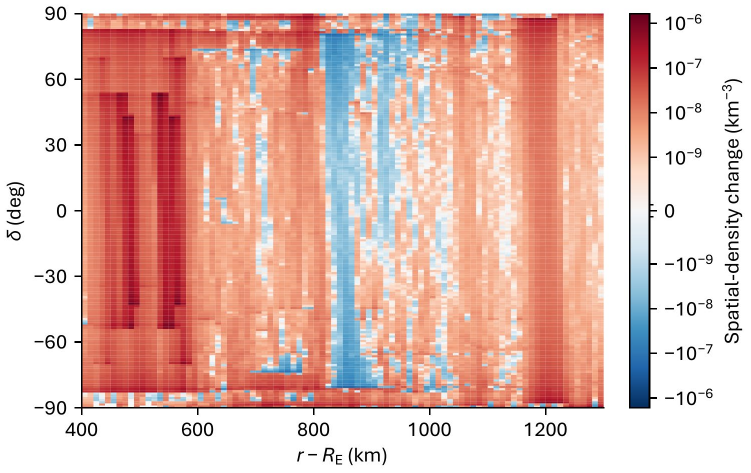}
    \caption{Change in spatial density from 2016 to 2025.}
    \label{fig:density_difference}
\end{figure}

Because $c$ is additive across catalog groups, the change for each target can be decomposed exactly within the adopted seven-group classification (Fig.~\ref{fig:target_decomposition}). For catalog group $g$ and target $q$, each heatmap cell reports the dimensionless contribution ratio
\begin{equation}
R_{q,g}=\frac{\Delta c_{q,g}}{c_q^{2016}},
\qquad
\Delta c_{q,g}=c_{q,g}^{2025}-c_{q,g}^{2016},
\end{equation}
where $c_q^{2016}$ and $c_q^{2025}$ are the target's total expected impact counts in 2016 and 2025, respectively, as defined in Eq.~\eqref{eq:mean_impact_count}. Positive and negative values indicate contributions that increase and decrease $c$, respectively.

The six targets exhibit different contribution patterns because their orbital residences intersect different portions of the changing catalog environment. For Circular 1, the annual expected impact count increases from $2.26\times10^{-6}$ in 2016 to $1.77\times10^{-5}$ in 2025, corresponding to a factor of 7.86 and a relative increase of 6.86 with respect to the 2016 value. The corresponding contribution ratios are 3.36 for Starlink, 2.22 for other payloads, 1.07 for unknown/TBA records, 0.14 for other debris, and 0.07 for rocket bodies. The named-debris-family contribution ratio is 0.003, while OneWeb contributes zero for this 500-km target under the reference cell setting, despite adding 654 accepted objects by 2025, because the target's orbital residence does not intersect the OneWeb altitude regime. The all-target decomposition extends beyond aggregate population-growth trends by quantifying how strongly each catalog group changes the target-specific metric relative to its 2016 reference value.

The two Korean assets also exhibit distinct contribution patterns. For KOMPSAT-3A, the dominant contributions arise from newly deployed Starlink satellites and other payloads, with contribution ratios of 7.58 and 3.76, respectively. In contrast, DOORY-SAT is comparatively less affected by changes in the catalog population; its largest contribution is from other debris, with a contribution ratio of only 0.22. These contrasting responses further illustrate that changes in the catalogued population affect individual space assets differently depending on their orbital regimes.

\begin{figure}[hbt!]
    \centering
    \includegraphics[page=13,width=0.98\linewidth]{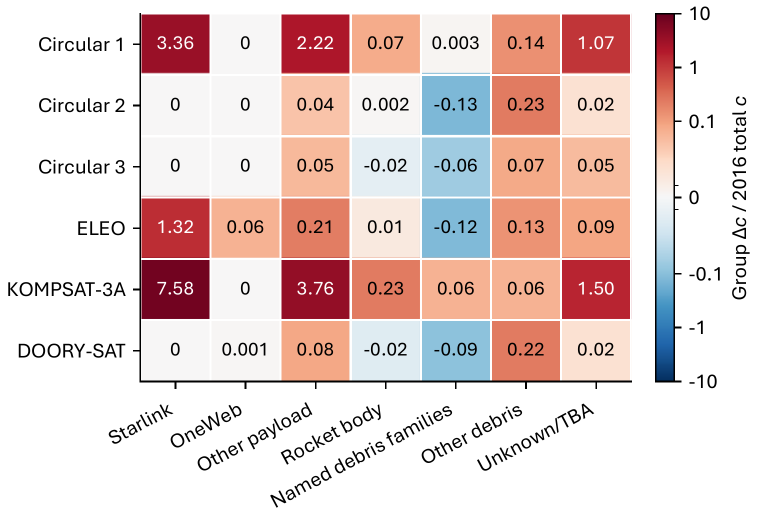}
    \caption{\rev{Catalog-group contributions to the 2016--2025 changes in expected impact count for the analysis targets, normalized by each target's 2016 value.}}
    \label{fig:target_decomposition}
\end{figure}

\subsection{Prospective Deployment: SpaceX's Orbital Data Center}

A prospective filing for SpaceX's ODC proposes four large orbital groups~\cite{spacex2026odc_response}. For the present analysis, the 94 proposed altitude shells are represented by one weighted circular orbit per shell, with the weight corresponding to the number of satellites assigned to that shell. Table~\ref{tab:odc_population} summarizes the modeled ODC population. The inclinations of the sun-synchronous shells are assigned according to their corresponding sun-synchronous conditions. The resulting modeled ODC population contains 998,240 satellites. The conditional stress test adds this population to the 2025 catalog environment.

\begin{table}[hbt!]
\caption{\rev{Modeled orbital groups for the ODC stress-test population.}}\label{tab:odc_population}
\rev{
\centering
\footnotesize

{\setlength{\tabcolsep}{4pt}\renewcommand{\arraystretch}{1.12}%
\begin{tabular*}{\textwidth}{@{\extracolsep{\fill}}lccccc@{}}
\toprule
\textbf{Group} &
\textbf{Inclination} &
\textbf{No. shells} &
\textbf{Altitude range (km)} &
\textbf{Satellites/shell} &
\textbf{Population} \\
\midrule
Lower $30^\circ$ & $30^\circ$ & 25 & 686--718 & 9,990 & 249,750 \\
Upper $30^\circ$ & $30^\circ$ & 25 & 946--978 & 9,990 & 249,750 \\
Lower SSO & SSO & 22 & 707--744 & 11,131 & 244,882 \\
Upper SSO & SSO & 22 & 967--1,002 & 11,539 & 253,858 \\
\midrule
\textbf{Total} & -- & \textbf{94} & -- & -- & \textbf{998,240} \\
\botrule
\end{tabular*}
}
}
\end{table}

The ODC stress test produces strongly target-dependent changes because only targets whose orbital residences intersect the added altitude regimes are affected. As shown in Table~\ref{tab:odc_results} and Fig.~\ref{fig:odc}, the annual probability of at least one impact increases from $1.52\times10^{-5}$ to $3.75\times10^{-3}$ for Circular 2 and from $4.47\times10^{-6}$ to $1.48\times10^{-3}$ for Circular 3. The ELEO target also shows a substantial increase, from $1.53\times10^{-5}$ to $2.66\times10^{-4}$, because its eccentric orbit spans multiple altitude regimes, including those populated in the ODC stress test. In contrast, Circular 1, KOMPSAT-3A, and DOORY-SAT show no change under the reference grid because their orbital residences do not intersect the added ODC shells.

Interestingly, Circular 2 and Circular 3 exhibit comparatively modest changes over the 2016--2025 historical snapshots, but become the two most strongly affected targets in the ODC stress test and show the highest annual impact probabilities among the six analyzed targets. This contrast illustrates how the concentration of a single large-scale constellation within particular orbital regimes can substantially alter the target-specific collision environment, even in regions that experienced relatively limited changes over the preceding decade.

\begin{table}[hbt!]
\caption{\rev{Conditional annual impact probabilities for the ODC stress test.}}\label{tab:odc_results}
\rev{
\centering
\small

{\setlength{\tabcolsep}{4pt}\renewcommand{\arraystretch}{1.12}%
\begin{tabular*}{\textwidth}{@{\extracolsep{\fill}}lcc@{}}
\toprule
\textbf{Target} &
\textbf{2025 $P_{\geq1}$} &
\textbf{ODC stress-test $P_{\geq1}$} \\
\midrule
Circular 1 & $1.7741\times10^{-5}$ & $1.7741\times10^{-5}$ \\
Circular 2 & $1.5170\times10^{-5}$ & $3.7485\times10^{-3}$ \\
Circular 3 & $4.4701\times10^{-6}$ & $1.4809\times10^{-3}$ \\
ELEO & $1.5331\times10^{-5}$ & $2.6631\times10^{-4}$ \\
\addlinespace[2pt]
KOMPSAT-3A & $1.9186\times10^{-5}$ & $1.9186\times10^{-5}$ \\
DOORY-SAT & $6.8746\times10^{-6}$ & $6.8746\times10^{-6}$ \\
\botrule
\end{tabular*}
}
}
\end{table}

\begin{figure}[hbt!]
    \centering
    \includegraphics[page=14,width=0.78\linewidth]{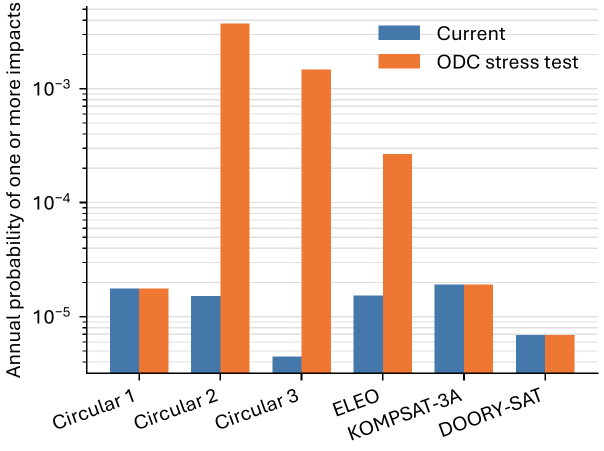}
    \caption{\rev{Annual impact probabilities for the 2025 catalog environment and the conditional ODC stress-test environment.}}
    \label{fig:odc}
\end{figure}

\subsection{Discussion}\label{sec:case_discussion}

The historical case study shows that catalog growth alone does not determine changes in target-specific collision metrics; the effect of each population depends on its spatial overlap with the target's orbital residence. The conditional ODC case shows the same behavior from a prospective perspective, with substantial changes concentrated among targets whose orbital residences intersect the added altitude regimes. Together, these case studies demonstrate the use of the 3D-cell workflow for catalog-conditioned attribution and comparative scenario analysis. The results should therefore be interpreted as macroscopic, catalog-conditioned assessments rather than endogenous population evolution or operational conjunction forecasts.

Several limitations define the scope of the present analysis. First, the adopted TLE inputs do not provide covariance information, and uncertainties associated with propagation from the selected TLE epochs to the common analysis epoch are not modeled. Second, the analysis represents only the supplied catalogued population and therefore excludes sub-catalog debris represented in engineering-environment models such as ORDEM and MASTER. Third, the collision metrics use prescribed effective area and relative speed rather than object-specific size, attitude, encounter geometry, and relative velocity. These limitations are consistent with the intermediate-detail role of the present model and motivate its complementary use with models operating at other levels of fidelity.

Accordingly, the 3D-cell workflow is intended to be used together with complementary models operating at different spatial, temporal, and decision scales. The present model provides rapid macroscopic characterization of a supplied orbital population, supports historical and scenario comparisons, and can identify targets or regions requiring further attention. Event-specific cases requiring uncertainty-aware collision assessment are more appropriately addressed using microscopic conjunction-analysis methods, while long-term changes driven by launches, disposal, breakup, drag, and collision feedback require evolutionary or source--sink models such as LEGEND and the MOCAT family. Within the integrated framework shown in Fig.~\ref{fig:integrated_context}, both a 1D source--sink model for long-term environment evolution and microscopic conjunction-analysis methods have been developed to complement the present 3D-cell workflow.

\section{Conclusions}\label{sec:conclusions}

Building on the previously reported 3D-cell formulation and implementation, this study establishes a reproducible and resolution-aware workflow for intermediate-detail macroscopic assessment of the orbital collision environment. The workflow retains individual catalog information and three-dimensional orbital residence while reducing encounter-level detail to support efficient environment screening and comparative analysis. The resolution study using the 2025 catalog and six representative targets, including Korean space assets, shows that target-specific expected impact counts can depend substantially and nonmonotonically on the selected cell widths, while the computational results demonstrate distinct runtime and memory trade-offs among the spatial dimensions. At the same time, the selected reference cell parameters provide a practical balance between computational cost and resolution sensitivity for the intended screening-level applications. These results emphasize the need to select spatial resolution according to the required level of detail and analysis scale.

The case studies demonstrate the use of the workflow for catalog-conditioned historical attribution and prospective scenario comparison. The 2016--2025 analysis shows that changes in target-specific collision metrics depend not only on overall catalog growth but also on the spatial overlap between individual populations and target orbits. The conditional ODC stress test similarly shows strongly target-dependent responses according to whether the added populations intersect the target's orbital residence.

The 3D-cell workflow is therefore intended as a complementary macroscopic tool for environment monitoring, comparative scenario assessment, and prioritization of cases requiring further analysis. Event-specific conjunction assessment requires higher-fidelity microscopic methods, while long-term population evolution requires evolutionary or source--sink models. Future work will focus on defining integrated analysis scenarios that connect these complementary model levels and demonstrating their combined use through representative SSA and policy-planning case studies. In parallel, adaptive or locally refined grids will be investigated to improve computational efficiency when higher spatial resolution is required.
}

\bmhead{Acknowledgements}

This research was supported by the Korea Astronomy and Space Science Institute under the Development of an Integrated Space Risk Response System (Project No. 2710103796), supervised by the Korea AeroSpace Administration \rev{(KASA)}.

\rev{
\bibliography{sn-bibliography}
}
\end{document}